\documentclass[]{gPAA2e}

\usepackage{xcolor}
\usepackage{soul}
\usepackage{array}

\begin{document}
\doi{10.1080/17445760.YYYY.CATSid}
 \issn{1744-5779}
 \issnp{1744-5760}
 \jvol{00} \jnum{00} \jyear{2011} 

\vspace{1in}
\begin{center}
{\Large Directing Chemotaxis-Based Spatial Self-Organization via Biased, Random Initial Conditions}
\vspace{2cm}
{\large \\ Sean Grimes, Linge Bai, Andrew W.E. McDonald, and David E. Breen$^*$}
\vspace{1cm}
\\ \{spg63,lb353,awm32,david\}@cs.drexel.edu\\
+1 215-895-2669
\vspace{0.3cm}
\\ Department of Computer Science\\
Drexel University\\
3141 Chestnut Street\\
Philadelphia, PA\hspace{3mm}19104\\
USA
\vspace{1in}
\\ This research was funded
by National Science Foundation grants CCF-0636323 and IIS-0845415.
\end{center}
\vspace{3.7in}
* Corresponding author.

\newpage

 \markboth{Sean Grimes, Linge Bai, Andrew W.E. McDonald, and David E. Breen}{Parallel, Emergent and Distributed Systems}


\title{Directing Chemotaxis-Based Spatial Self-Organization via Biased, Random Initial Conditions}
\author{Sean Grimes, Linge Bai, Andrew W.E. McDonald, and David E. Breen$^{\ast}$\thanks{$^\ast$Corresponding author. Email: david@cs.drexel.edu, +1 (215) 895-1626 \vspace{6pt}}  \\\vspace{6pt}  \em{Department of Computer Science, Drexel University  Philadelphia, PA   19104  USA}}

\maketitle

\begin{abstract}
Inspired by the chemotaxis interaction of living cells, we have developed
an agent-based approach for self-organizing shape formation. Since all
our simulations begin with a different uniform random configuration
and our agents
move stochastically, it has been observed that the self-organization
process may form two or more stable final configurations. These differing
configurations may be characterized via statistical moments of the
agents' locations. In order to direct the agents to robustly form one
specific configuration, we generate biased initial conditions whose
statistical moments are related to moments of the desired configuration.
With this approach, we are able to successfully direct the aggregating
swarms to produced a desired macroscopic shape, starting from randomized
initial conditions with controlled statistical properties. 
\end{abstract}

\bigskip

\section{Introduction}

\begin{figure}[h!]
\begin{center}
\includegraphics[width=\linewidth]{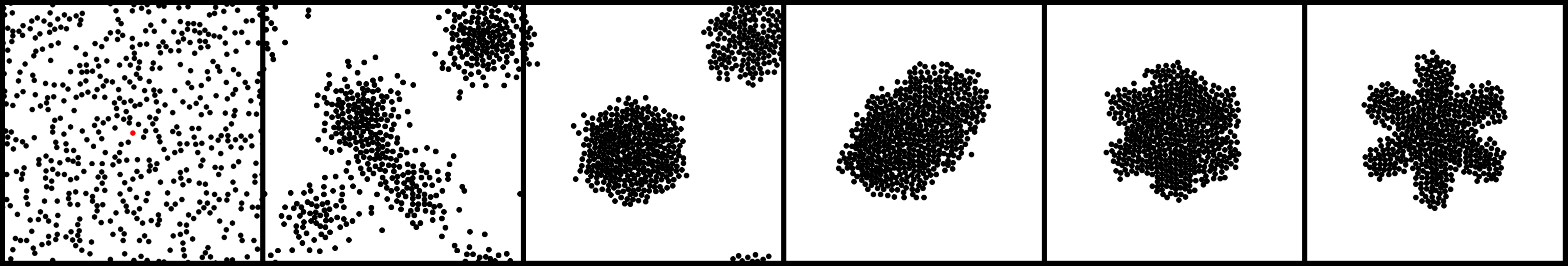}
\end{center}
\caption{Morphogenetic Primitives self-organizing into a star shape.
Initially published in \cite{BaiChapter2012}.}
\label{fig:star}
\end{figure}

Motivated by the ability of cells to form into specific shapes and
structures, in previous work we developed chemotaxis-inspired
software agents for
self-organizing shape formation \cite{Bai08b,BaiChapter2012}.
The actions of the agents, which we call Morphogenetic Primitives (MPs),
are based on
the behaviors exhibited by living cells. Cells emit chemicals into the
environment. Neighboring cells detect the overall chemical concentration at
their surfaces and respond to the chemical stimulus by moving along the
chemical field's gradients~\cite{eisenbach2004}.
Similarly, in our system the agents emit
a virtual chemical, with its concentration 
defined by an explicit mathematical expression. 
A set of agents start with an initial random configuration and
stochastically
follow the gradient of the cumulative concentration field.
These chemotaxis-based local interactions direct the agents to 
self-organize into user-specified shapes (Figure \ref{fig:star}).
Since the behaviors of MPs are based on local information and interactions,
they could provide a distributed, scalable approach for controlling the
movements of a robotic swarm.

In some cases, we
have observed though that the agents do not spatially self-organize into 
a unique shape, but instead form
two or more stable final configurations.
If MPs are used to control the motions of individual robots,
it would be extremely useful to direct the outcome of
these bifurcating spatial self-organization processes towards a
consistent outcome.
This would allow us to guarantee that all of our MP aggregations
produce a single, desired shape.  This is a property that is essential
for robust and predictable swarm control algorithms, one that would make
the algorithm reliable for engineering applications.
Towards this end, we have analyzed whole swarm populations at a global
level in search of macroscopic, distinguishing attributes.
This analysis identified features, based on statistical moments of
the agents' positions, that have significantly different values
for different outcomes of a swarm aggregation.
In earlier work we discovered that these statistical moments can
be used to accurately
predict the outcome of the self-organization process at an early stage
of the shape aggregation~\cite{bai2014scw}.
Given these differentiating moments, the work described here
investigated techniques for directing the outcome of our
self-organizing system
via biased, random initial conditions in order to consistently
produce a desired final configuration.

Through our study of the dynamics of a swarm's statistical moments
during the aggregation process we noted the connection
between initial conditions and the final shape configuration 
of the swarm.  We discovered that 
biased, random initial conditions that meet specified constraints,
i.e.\ have well-defined statistical properties,
robustly yield simulations with a unique final outcome.
For those agent interactions that ultimately
produce bifurcating/multiple shapes,
we have identified for each shape, 
the most distinguishing macroscopic statistical moment
of the evolving swarm.
It is possible to generate random distributions of MPs
that have specific statistical moments.
Our work empirically shows that for bifurcating self-organizing,
non-linear, dynamical systems (e.g. a swarm of Morphogenetic
Primitives)
one final outcome can be consistently generated
by enforcing a constraint on the value of a single moment
when generating the swarm's initial conditions.
Given this feature of our system, we are able to control the final
outcome of the simulation by simply
thresholding the value of a statistical moment for a particular
starting distribution, i.e., we constrain the random initial conditions
to have specific statistical properties.

\section{Related Work}

Research on distributed agent-based systems that can form spatial
patterns and shapes, as well as swarm behaviors, has been conducted
for several decades. Reynolds \cite{Reynolds87} proposed the
seminal model for simulating flocking and schooling behaviors based
on the local interactions of ``boids''.
Fleischer and Barr \cite{Fleischer94,Fleischer96} explored a cell-based
developmental model for self-organizing geometric structures.
Theraulaz and Bonabeau \cite{Theraulaz95b,Theraulaz95a} presented a
modeling approach based on the swarming behavior of social insects.
They combined swarm techniques with 3D cellular automata to create
autonomous agents that indirectly interact in order to create complex
3D structures. Viscek et al.~\cite{Vicsek95} investigated a particle-based
model related to the Reynolds model and found that macroscopic
phase changes occurred in the particle system when introducing noise
in the local interactions. Jadbabaie et al.~\cite{Jadbabaie03}
further explored the Vicsek model and provided a theoretical explanation
for the model's observed behavior, as well as convergence results for
classes of switching signals and arbitrary initial heading vectors.

The initial work in this area of research created distributed,
locally-interacting agent-based systems, then observed and 
characterized their behaviors.  Later work explored techniques for
directing these distributed, self-organizing systems.
Eggenberger Hotz \cite{eggenberger1997ems,hotz:cdp} proposed the
use of genetic regulatory
networks coupled with developmental processes for use in artificial
evolution and was able to evolve simple shapes.
Bonabeau et al.~\cite{Bonabeau00} applied genetic algorithms
to the stigmergic swarm-based 3D
construction method of Theraulaz and Bonabeau in order to
evolve interactions that produce user-acceptable structures.
Nagpal et al.~\cite{nagpal2002psa,Nagpal:2003:PMB} presented
techniques to achieve
programmable self-assembly. Cells are identically-programmed units which
are randomly distributed and communicate with each other within a local
area. In this approach, global-to-local compilation is used to generate
the program executed by each cell, which has specialized initial
parameters.  Stoy and Nagpal \cite{stoy2004sru} presented an approach
to self-reconfiguration based on directed growth, where the desired
configuration (which is stored in each module) is grown from an initial
seed module.  Spare modules move along recruitment gradients emanating
from attached modules to create the final shape. Gradients derived
from global potential fields have also been investigated for
directing robot swarms. Both Rimon and Koditschek \cite{Rimon1992} and
Hsieh and Kumar \cite{MH:06}
demonstrated that robot paths and controls can be computed from
these fields, which lead the robots to form a pre-defined shape.

Shen et al.~\cite{shen2004hormone} proposed a Digital Hormone Model
for directing
robot swarms to perform such tasks as surrounding a target, covering
an area and bypassing barriers. The model relies on local communications
between identical agents, but it also has each agent move
towards a single global target.
Swarm chemistry, proposed by
Sayama \cite{sayama2009swarm,sayama2010robust} and based on
Reynolds' model, is an approach for designing spatio-temporal patterns
for kinetically interacting, heterogeneous agents. An interactive
evolutionary method, similar to Sims' \cite{Sims92}, has been used
to define system parameters that lead to agent segregation and structure
formation.  Mamei et al.~\cite{mamei2004experiments} proposed a
distributed algorithm for
robots that are attracted to and aggregate around targets sensed over short
distances. By electing leader(s) as
barycenter(s), propagating gradients of varying structure and using these
gradients as instruction conditionals, a swarm of simulated robots
are able to self-organize into a number of simple shapes such as a circle,
ring, and lobes.
Von Mammen and Christian \cite{von2009evolution} described
swarm grammars, an agent-based extension of Lindenmayer systems, that
are capable of adapting to their environment and evolve agent parameters
in order to create structures that incorporate aspects of developmental
design and morphogenesis. The field of Guided Self-Organization (GSO)
\cite{ProkopenkoBook01} has developed techniques for steering
self-organizing systems towards desired
outcomes, while still attempting to not constrain the
system's configuration space during its evolution.

Doursat \cite{Doursat2008GCB,Doursat2008OGA}
proposed a model for artificial development which combines proliferation,
differentiation, self-assembly, pattern formation and genetic regulation.
Via genetic-like regulation at the agent level, the agents can
self-organize into a number of patterned shapes and structures.
Werfel et al.~\cite{werfel2014designing} proposed a
decentralized multi-agent system approach, inspired by mound-building
termites, for building user-defined structures. A user specifies a
desired structure, and the system automatically generates
low-level rules for independent climbing robots that guarantee
production of the structure.
A single ``seed'' brick is used as a landmark to identify where the
structure is going to be built, and defines the origin of a shared
coordinate system for the robots.
Gerling and Von Mammen \cite{gerling2016robotics} provided a context
for this type of work in a summary of self-organized approaches to
construction.

Our previous and latest work on agent-based shape formation stands apart
from related work in that it
utilizes a chemotaxis-based interaction paradigm inspired by the
behaviors of living cells which leads to the formation of tissues.
Given the goal of recreating the properties of cells, MPs
were designed with principles that should make them scalable and
robust \cite{BaiChapter2012}. These design principles define MPs as
identical, distributed agents that are not directed by a 'master designer',
exchange information locally, carry no representation of the shape to
be formed, and have no information about their global location. 
The macroscopic shape of the swarm emerges from the aggregation of local
interactions and behaviors. 
Our approach is therefore novel compared to previous work in that it
contains all of the following features. 1) All morphogenetic primitives
are randomly placed in the environment, are identical, and perform the
same simple actions, unlike Nagpal et al.~\cite{nagpal2002psa,Nagpal:2003:PMB}),
Mamei et al.~\cite{mamei2004experiments}, and
Doursat \cite{Doursat2008GCB,Doursat2008OGA}.
They require no differentiated behaviors or customized initialized states.
2) No initialization of spatial information is needed in the
computational environment, unlike Stoy and Nagpal \cite{stoy2004sru},
Shen et al.~\cite{shen2004hormone}, and Werfel et
al.~\cite{werfel2014designing}.
3) Individual MPs do not know their location in any external/global
coordinate system, unlike Stoy and Nagpal \cite{stoy2004sru},
and \cite{werfel2014designing}.
4) MPs do not contain or utilize a representation of the predefined
global shape that is being composed, unlike Stoy and Nagpal \cite{stoy2004sru},
Rimon and Koditschek \cite{Rimon1992}, and Hsieh and Kumar \cite{MH:06}.
5) We 
utilize genetic programming to discover the MP concentration field
functions that lead to the formation of a user-specified shape.
Chemotaxis then
provides a straightforward mechanism for determining the motion of
MPs, in contrast to the difficult-to-program approaches
of Shen et al.~\cite{shen2004hormone} and
Sayama \cite{sayama2009swarm,sayama2010robust}.
Note that our new work described here enhances our previously developed
system \cite{BaiChapter2012} 
to make MPs more robust, consistent and reliable.

\section{Background Material}

\subsection{Agent-based Shape Formation}

Like Pfeifer et al.~\cite{pfeifer2007self} we turn to biology and
self-organization for
insights into the design of autonomous robots, robotic swarms in our case. 
Our previous work in self-organizing shape
formation~\cite{BaiChapter2012,Bai13} is inspired by
developmental biology~\cite{GilbertBook2013} and
morphogenesis~\cite{DaviesBook}, and
builds upon a chemotaxis-based cell aggregation
simulation system~\cite{Eyiyurekli08a}.
Morphogenesis is the process that forms the shape or structure 
of an organism through cell shape change, movement, attachment,
growth and death. 
We have explored chemotaxis as a paradigm for agent system control
because the motions induced by chemotaxis (one of the mechanisms of
morphogenesis) may produce patterns, structures or sorting 
of cells~\cite{SekimuraBook}.

\subsubsection{Morphogenetic Primitives}

Morphogenetic Primitives are initially
placed inside a $2D$ environment with a random uniform
distribution.
Each MP is represented by a small disc and emits a `chemical' 
into the environment within a fixed distance relative to its own
local coordinate system. Every MP emits
the identical local chemical field.
An MP
detects the cumulative chemical field at eight receptors on its
surface, and calculates
the field gradient from this input. MPs move in the direction of the
field gradient with a speed proportional to the magnitude of the gradient. 
By employing these relatively simple chemotaxis-inspired
behaviors MPs are able to self-organize into user-specified
macroscopic shapes.

\begin{figure}
\begin{center}
\includegraphics[width=\linewidth]{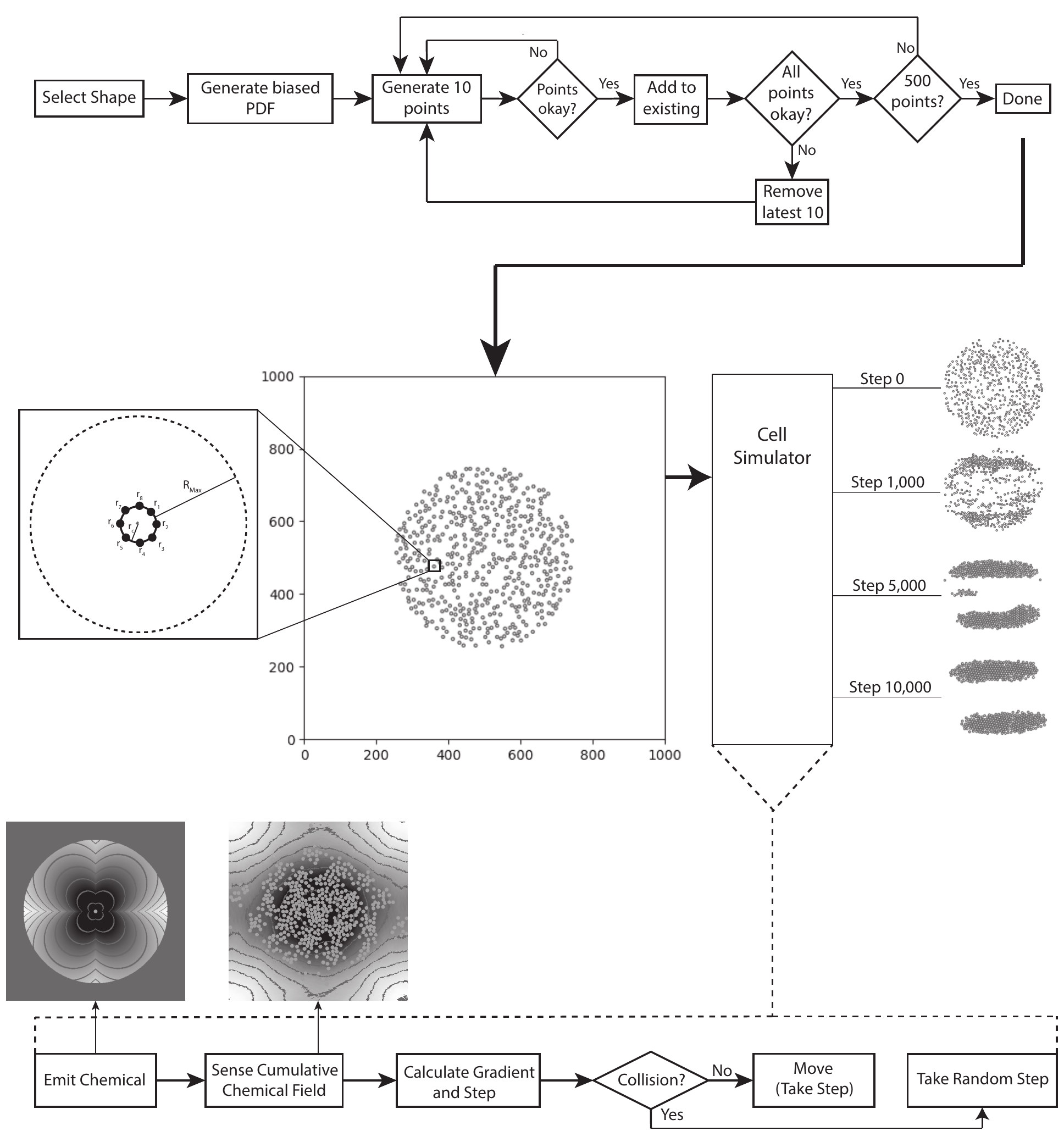}
\end{center}
\caption{Schematic diagram of the directed self-organization process
based on specifying biased initial conditions for 
Morphogenetic Primitives.}
\label{fig:schematic}
\end{figure}

This process is schematically presented in the bottom 2/3 of Figure
\ref{fig:schematic}. In the middle left of the figure, a close-up
of an MP is provided showing its eight chemical sensors and the range of
the finite chemical field that it emits. Numerous MPs are randomly placed
in the computational arena, and are provided as initial conditions to a
chemotaxis-based cell aggregation simulator. The simulator then computes
an aggregation simulation based on the one chemical field that is
associated with all MPs. The self-organization of the agent swarm is
shown in the middle right of the figure. The bottom flowchart of the
figure outlines the steps taken by the cell simulator for each cell.
The bottom left image shows a representative MP chemical field, with the
chemical concentration visualized with gray-scale colors. Isolines
are added to highlight the structure of the chemical field. The image
to its right shows a cumulative chemical field given the contributions
of all of the MPs in the arena. 
The top part of the figure will
be explained later in the paper.

\subsubsection{Genetic Programming for Discovering Local Interactions}

While MPs' fundamental interactions are
based on a chemotaxis-inspired paradigm,
we do not limit their behaviors/properties to be physically realistic or
completely consistent with biology. Instead, developmental biology provides a
motivating starting point for MPs. As a way to customize chemotaxis-inspired
agents for shape formation, we alter the chemical concentration
fields around
individual cells. Instead of the chemical concentration dropping off 
only as an inverse
function of distance $d$ from the cell's surface (e.g. $1/d$),
in our system we define the concentration field
with an explicit function of $d$ and $\theta$, the angular location
in the cell's local coordinate system.

\begin{figure}
\begin{center}
\includegraphics[width=\linewidth]{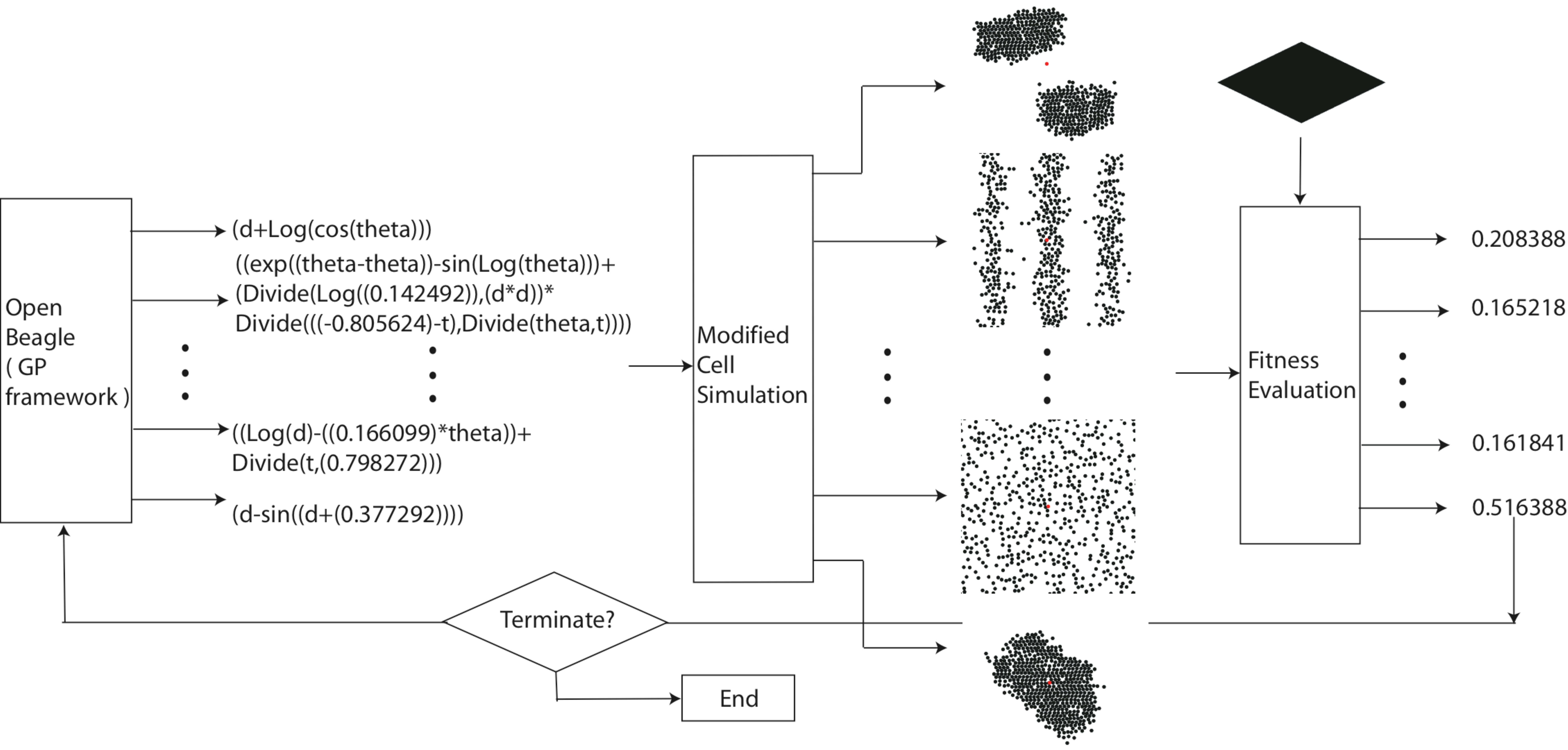}
\end{center}
\caption{The genetic programming process that produces the local
chemical field functions of the shape primitives. Initially published
in \cite{Bai08b}.}
\label{fig:flowchart}
\end{figure}

Currently, there is no prescriptive way to specify a particular local
field function that will direct MPs to form a specific macroscopic shape,
we therefore
employ genetic programming~\cite{Koza92Book} to produce the 
mathematical expression that
explicitly specifies the field function. In order to meet the substantial
computational requirement imposed by our evolutionary computing approach, we
have implemented a master-slave form of the distributed genetic
programming process \cite{Bai08b}. The fitness measure associated with
each individual field function is
based on the shape that emerges from the chemical-field-driven aggregation
simulation, and determines which functions will be passed along to later
generations. The genetic process stops once an individual (i.e., a mathematical
expression) in the population produces the desired shape via a chemotaxis
simulation, or after a certain number of generations have been produced and
evaluated. Figure~\ref{fig:flowchart} illustrates this approach. See
\cite{Bai08b}, \cite{BaiChapter2012} and \cite{LingeMSThesis} for more
details on MPs and
the software system that implements them.

With this algorithm, we have successfully evolved local MP chemical field
functions for a number of simple shapes~\cite{BaiChapter2012}.
These results support the proposition that biological phenomena offer
paradigms for designing cellular
primitives for self-organizing shape formation.
While the resulting
explicit chemical fields are not biologically/chemically plausible,
they do provide an approach for controlling robot swarms that communicate
wirelessly over short distances and share minimal information with each
other. Thus the agents in the swarm do not require significant compute
power to self-organize.
Additionally, evolutionary
computing techniques, specifically genetic programming, have been crucial
for discovering the detailed local interactions that lead to the
emergence of the swarm's macroscopic structure.

However, given the MPs' initial random
configurations and the stochastic nature of the self-organization process,
the outcomes of the simulations with a specific field function are not
always the same. We have found
that the shape formation simulations, which include random
displacements of the MPs and noise in their movements, can generate
bifurcating results. For some field functions, if we
run numerous simulations each starting with a different random uniform
distribution of MPs, two sets of final configurations will be formed.
In most cases an equal number of each configuration are produced, but
in a few cases the ratio of the numbers is not one.
Since it would be useful to control the outcomes of the self-organization
process, we have developed methods for directing the final configuration
of a bifurcating simulation by starting the simulation
with biased initial conditions \cite{LingePhDThesis}.
 
\subsection{Outcome Prediction}

The first step towards developing methods that direct the outcome
of a swarm simulation involved identifying spatial features that are
correlated
with and can differentiate the final, different swarm configurations. 
Our initial effort towards achieving this goal investigated
methods for predicting the final configuration of a bifurcating simulation
at an early stage of the aggregation
process.
Our reasoning was that if certain spatial features can be used
to predict the outcome of an aggregation, then they represent unique
attributes of the swarm that could be manipulated to direct
the swarm.
In order to predict the final outcome of a self-organizing shape formation
simulation, we first extracted features that capture the
spatial distribution of the MPs. Moments provide a quantitative way to
describe a distribution. Since MPs are defined as small discs, we
use the center of each disc to represent each MP's location. We
therefore can simplify the collection of MP locations
as a set of $2D$ points, and apply moment analysis to this set
over the duration of the MP simulation.

We calculated the mean (first moment), variance (second central moment),
skewness (third central moment) and kurtosis (fourth central moment)
from the $x$ and $y$
coordinates of the MP centers. We analyzed the locations $X_i$ of all
points (MPs) as a whole, rather than tracking the location and movement
of each individual point.
The population size of the agents is denoted as $n$, $(n=500)$, and the
formulas of the four moments $M_1$ to $M_4$ are given in
Equations~\ref{eq:moment1} to~\ref{eq:moment4},
\begin{align}
M_1 &= \frac{1}{n} \sum_{i=1}^n X_i, 
\label{eq:moment1} \\
M_2 &= \frac{1}{n} \sum_{i=1}^n (X_i - M_1)^2,
\label{eq:moment2} \\
M_3 &= [\frac{1}{n} \sum_{i=1}^n (X_i - M_1)^3] / (M_2)^{3/2}, 
\label{eq:moment3} \\
M_4 &= [\frac{1}{n} \sum_{i=1}^n (X_i - M_1)^4] / (M_2)^2.
\label{eq:moment4}
\end{align}
 
These statistical moments provide quantitative information about the
shape of histograms/distributions. When computed for the $x$ and $y$
coordinates of the MPs, these moments capture the asymmetry and shape
of the spatial distribution of the whole population. 
We have not found
it necessary to compute cross moments, with the first four
moments providing sufficient information for our analysis.
Since the $x$ and $y$ coordinates of the points change over time,
so do the four
moments of the distribution of the $x$ and $y$ values.
The change of the moments as a function of simulation time also
provides insight into the dynamic nature of a particular MP
simulation.


At each simulation time $t$, the four moments $M_i(t)$ ($i=1~to~4$)
of the overall distribution are calculated.
We then approximate the time derivative of the moments as the slope of
a linear interpolating function of consecutive moment values.
By calculating the moments and their time derivatives
for both the $x$ and $y$ coordinates of the point set, at a given time $t$,
we obtain a 16-dimensional vector to represent the distribution,
\begin{flushleft}
$M_{x_1}(t), M_{y_1}(t), M_{x_2}(t), M_{y_2}(t),
M_{x_3}(t), M_{y_3}(t), M_{x_4}(t), M_{y_4}(t),
k_{x_1}(t), k_{y_1}(t),$
\end{flushleft}
\begin{flushright}
$k_{x_2}(t), k_{y_2}(t),
k_{x_3}(t), k_{y_3}(t), k_{x_4}(t), k_{y_4}(t)$.
\label{expr:featurevector}
\end{flushright}

Given the sensitivity of non-linear dynamical systems to initial
conditions \cite{WigginsBook}, it makes it extremely difficult, if
not impossible, to predict the outcome of our complex, self-organizing
system from its initial, random spatial configuration.
We therefore attempted to predict the final spatial configuration
at an early stage of the aggregation, usually before it is visually
evident what shape will emerge from the process.
We considered prediction of the bifurcating outcomes as a classification
problem and utilized support vector machines (SVMs)~\cite{SVMBook} to
solve it.  We have found that
applying SVMs to the distribution feature vector at a simulation time
that is a
small percentage of the total time needed for the final aggregated
shape to form produced acceptable results.
Given 200 MP simulations for a variety of bifurcating self-organizing
shapes we found that we could predict the outcome of the aggregation
at a time point 5\% to 10\% into the simulation with an accuracy
of 81\% to 91\%.  We view these results as satisfactory because they
demonstrate that a strong correlation between a swarm's moments
and its final formed shape does exist.  More details about
this study may be found in \cite{LingePhDThesis} and \cite{bai2014scw}.

\section{Directing Spatial Self-Organization}

Since the outcome of an MP simulation can be predicted
at an early stage of aggregation using the moments of the agents'
positions, we then explored methods for controlling the swarm via
manipulating the moments of the swarm's initial configuration.
The general strategy is to create random initial configurations
for the MP simulations, but with constrained, biased moments.
We have observed that this strategy can consistently direct the
swarm to aggregate
into specific final configurations.  The first step of the strategy
analyzes the bifurcating simulations to determine which of
the moments diverge the most for the two different outcomes. This is the 
moment that will be biased in the swarm's initial, random 
configuration.

\subsection{Moment Analysis}

\begin{figure}
\begin{center}
\includegraphics[width=1.0\linewidth]{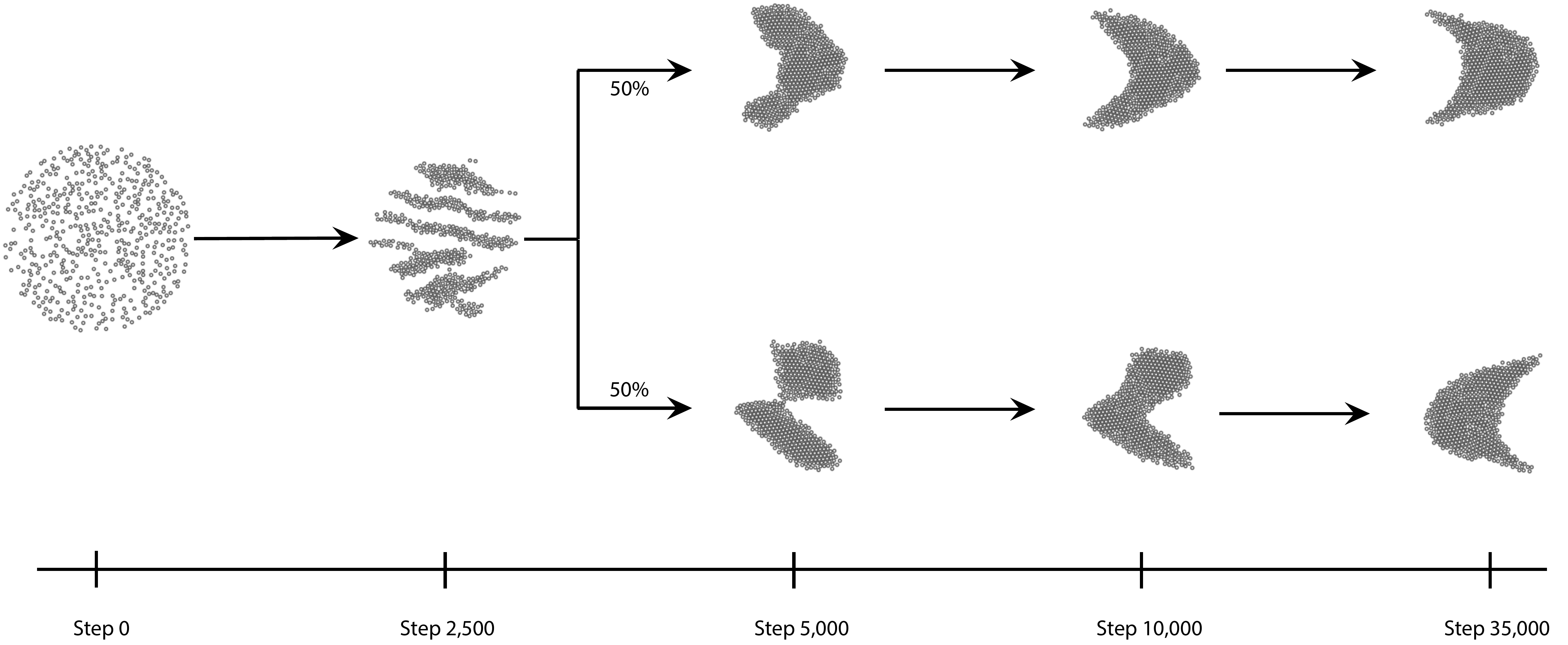}
\end{center}
\caption{Shape aggregation of the quarter-moon MPs starting
from random initial conditions.}
\label{fig:quartermoon}
\end{figure}

\begin{figure}
\begin{center}
\includegraphics[width=0.9\linewidth]{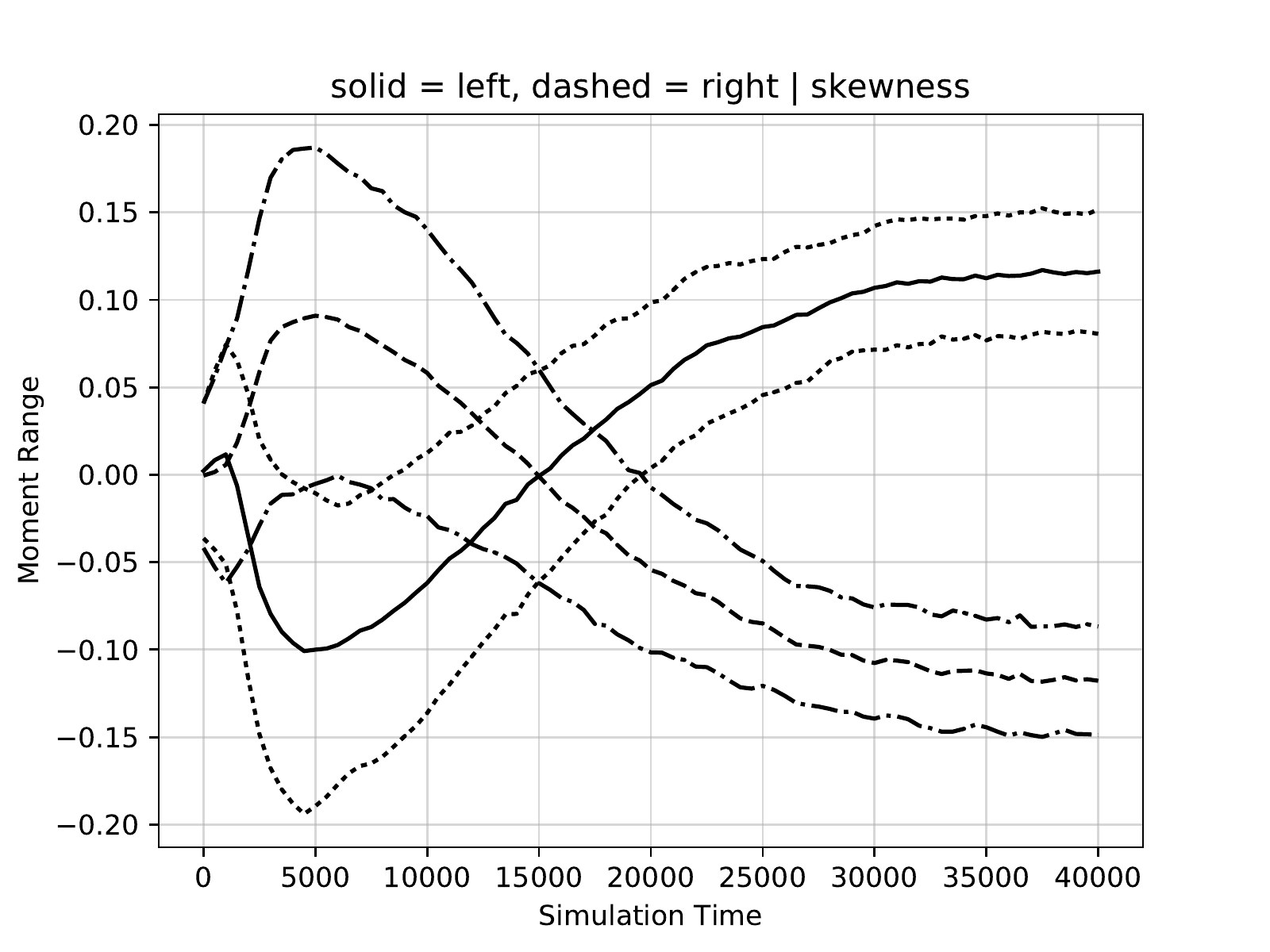}
\end{center}
\caption{Skewness of the $x$ coordinate of the unbiased quarter-moon shape
aggregations over time.}
\label{fig:quartermoon_x3}
\end{figure}

One of our aggregations, which produces what we call the quarter-moon shape,
provides an example of the moment analysis.
Of the $200$ simulations starting with
a uniform random, unbiased initial condition,
$100$ produce left-pointing structures and $100$ produce right-pointing 
structures.
Figure~\ref{fig:quartermoon} shows a typical swarm aggregation for this
shape.  
The four moments of both $x$ and $y$ coordinates are calculated over
all 35,000 simulation steps. Additionally the mean and standard
deviation of each statistical moment are calculated
for the two categories, i.e.\ left-pointing and right-pointing,
over the simulation time steps. Plotting the mean and the
mean $\pm$ standard deviation of the moments over time immediately
highlights the moments which are the most differentiating and may
be used to identify specific shapes.
For the quarter-moon example the third $x$ moment (skewness) is the one
with the greatest separation of values for the two possible outcomes,
as seen in Figure~\ref{fig:quartermoon_x3}. The solid and dashed lines
are the mean of the skewness of the $x$ coordinate for the two outcomes.
The dotted and dot-dashed lines are
mean $\pm$ standard deviation. The dashed curve is produced from
structures that are
right-pointing and follow the path in the top of
Figure~\ref{fig:quartermoon}. The solid curve is produced from
the left-pointing
structures, with a typical aggregation presented at the bottom of
Figure~\ref{fig:quartermoon}. 

By analyzing the time series in Figure~\ref{fig:quartermoon_x3}, we see
that the skewness of the $x$ coordinates of the two classes
starts at about the same value, approximately $-0.05$ to $0.05$, at time
step $0$.  The values should be near zero, since all simulations
begin with uniform random configurations.
The skewness of the two classes first separates by increasing or decreasing,
followed by a zero crossing and then a reversed trend appears
until they reach their final states at step
35,000. Observing the values for the solid and dashed curves over all
simulation time steps, we can identify three regions in the plot: a
region occupied by solid/dotted curves only, a region occupied by
dashed/dot-dashed curves only
and an overlapping region.
To be specific, considering the values of $x$ skewness in
Figure~\ref{fig:quartermoon_x3} (by projecting the curves into
the $y$ axis),   
the range of $[-0.195, -0.150]$ is covered by solid/dotted curves
only; $[0.150, 0.190]$ is covered by dashed/dot-dashed curves only and
$[-0.150, 0.150]$ is covered by both outcomes. 

When determining the appropriate threshold value for a constrained
moment, we start with the mean value of the non-overlapping
moment range for a particular shape, and then adjust if needed.
For the region covered
only by the quarter-moon's dashed/dot-dashed curves (for right-pointing
shapes) the mean value of skewness is $0.170$. While using this
value for the
moment constraint produced reasonable results ($94\%$ of the biased
initial conditions produced the desired shape), we found, via multiple
experimental runs, that the
threshold value on the skewness had to be increased to produce
a consistent result. In general this was the process employed for
determining the
moment constraint thresholds needed to generate the desired outcomes.

\subsection{Generating Constrained Biased Distributions}

Once the most significant distinguishing moment (the one with the greatest
value difference in the final configuration) for a shape is identified,
the information is utilized to direct the shape aggregation by imposing
constraints on this moment in the initial conditions.
We assume that the MPs' $x$ and $y$ coordinates are independent,
and therefore create two probability density functions, each
representing the $x$ and $y$ coordinate.
One probability density function is
created for the constrained coordinate and the other coordinate is
considered to be uniformly random.
Samples are drawn independently from the two
distributions to produce a single $(x,y)$ location.

This approach generates random distributions, i.e.\ 2D random initial
configurations for the shape simulation, that meet a constraint
on a particular moment in the $x$ or $y$ coordinate. We have found
that constraining one of the eight moments (mean, variance, skewness and
kurtosis for $x$ and $y$) is sufficient for producing satisfactory
results. 
Constraining multiple moments does not significantly improve the outcomes,
and would further complicate the process of generating initial conditions.
Once one significant moment of a distribution and its
threshold value have been identified, the remaining three moments
for that spatial coordinate ($x$ or $y$)
may be set to values observed in uniform random distributions.
These values are $mean = 500$ (the center of our computational arena),
$variance = {15,000}$, $skewness = 0$ (to make the distribution symmetric),
and $kurtosis = 2$.
Once the four moments for one of the coordinates have
been specified, a probability density function (PDF) with those moments is
defined.  The values for the other coordinate are generated from a
uniform random distribution.

For the constrained dimension we create a Gram-Charlier expansion of
the normal
distribution (chosen for its convergence properties) with specified
moments \cite{seabold2010statsmodels}.
This Gaussian-expanded probability
density function is then discretized with 100,000 samples with values
falling within the range of 0 to 1000 (the range of the
computational arena).  Slice sampling~\cite{neal2003slice}, a Markov chain
sampling method chosen for its efficiency, 
is then utilized to draw 500 samples from this
discretized distribution. 
Theoretically, the specified PDF can be sampled to
produce a distribution that has the same moments as the PDF. Our
experience has shown that the sample size needs to be quite large
(on the order of 1 million) for this to be true.  For our sample size
of 500 (the number of MPs in an aggregation simulation) and heavily biased
PDFs, the resulting moments of the finite sample set do not necessarily
match the ones desired for a particular shape.

As the distribution generation process cannot guarantee that a
sampled distribution will meet the required moment constraints,
the four moments of the generated distributions for the
constrained coordinate
are computed to determine if the constraint is actually met.
If the value of the significant moment is not above or below 
(depending on the constraint to be enforced) the
specified threshold, the sampled distribution is rejected.
If the constraint is met, the distribution is accepted as the
initial conditions for a simulation computation.


Since our agents are not points, but in fact are
discs with a fixed radius, we maintain a distance of $2R$ (where
$R$ is the radius of a disc) between sample points, to ensure that
the MPs do not overlap. Therefore, if an $(x,y)$ pair is generated that is
less than $2R$ distance to a previously generated location, it is
rejected and another $(x,y)$ pair is calculated.  This process
continues until a sufficient number of MP locations are generated
for the initial conditions.

We have found that it is more computationally efficient to generate
numerous smaller sample sets and then merge them into a single
point set, rather than attempt to compute a single, large point
sample, when composing biased initial conditions for our aggregation
simulations.  We apply this approach by drawing 50 subsets,
$T_{k = 1,...,50 }$ of 10 samples,
rather than 1 set of 500 samples.
Each sample drawn is checked for overlap with existing samples, discarded if overlap exists,
and otherwise added to the current $T_k$.
Once each $T_k$ has 10 samples, it is checked for conformance to
the moment restrictions prior to insertion into the
final set, $S$, of 500 samples. If a given $T_k$ fails, a new $T_k$
is drawn. An acceptable $T_k$
is merged with $S$, which is then checked for conformance to the
moment restrictions.
If the updated $S$ does not pass, it is reverted to its prior state, $S - T_k$, and a new $T_k$ is drawn.
This process continues until $|S| = 500$.
The algorithm for generating biased initial condition, once the initial
sampling does not meet the moment requirements, is diagrammed in the
flowchart at the top of Figure \ref{fig:schematic}.
Via this approach, we are able to generate initial conditions for 
our computational experiments in a few seconds, as
opposed to several hours, when attempting to generate all 500 points
of $S$ at once for certain ``extreme'' biased conditions, e.g.~low
kurtosis.



\section{Results}



\begin{figure}
\begin{center}
\hspace{1.5mm}
\includegraphics[width=1.00\linewidth]{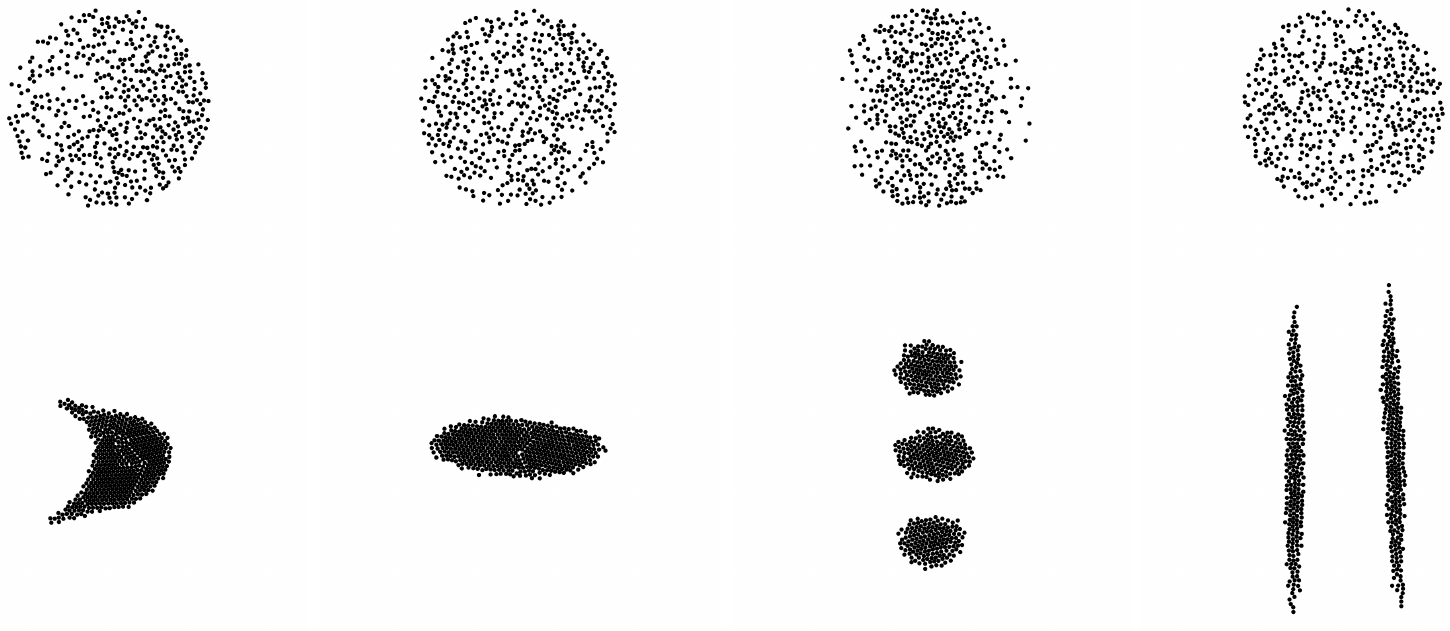}\\
\end{center}
\vspace{-2mm}
\hspace{9mm}(a)\hspace{33mm}(b)\hspace{35mm}(c)\hspace{34mm}(d)
\vspace{3mm}
\caption{(top row) Biased initial conditions ((a) $x$ skewness = -0.315, 
(b) $y$ kurtosis = 2.150, (c) $x$ variance = 9,596,
(d) $x$ kurtosis = 1.88) that robustly evolve (bottom row) into (a) a
right-pointing quarter-moon,  (b) a single
ellipse, (c) three discs and (d) two line segments.}
\label{fig:biased_results}
\end{figure}

We have applied our method for directing spatial self-organizations,
which generates biased initial 
configurations, to a number of bifurcating shape
aggregations.
We refer to the resulting shapes as 
the quarter-moon, ellipse,
discs, and two parallel line segments.
These shapes (and their associated chemical fields) were utilized in an
earlier study \cite{bai2014scw},
and had shown not to produce a single final, aggregated result.
In our initial MP work it was not uncommon for the output of the evolution
process to generate chemical fields that led to the formation of
different shapes from a single field. These four were chosen
because they generated two different shapes in equal proportions (except
for the parallel lines shape) from uniformly random initial conditions.
The chemical field functions that direct MPs to form into these
shapes are detailed in Table \ref{table:functions}.
Given biased initial conditions the aggregation simulations
produce one final outcome in almost all of cases.
Moreover, by thresholding the moment constraints on the
biased initial conditions it is possible to control which shape is
produced by a simulation.
Figure~\ref{fig:biased_results} (top row) shows biased initial conditions
for a number of shape aggregations created with this method.
The bottom row illustrates the final outcome of each MP simulation
that is produced from the associated biased starting configuration. 

{\renewcommand{\arraystretch}{1.2}
\begin{table}[t]
\begin{center}
\begin{tabular}[t]{| c | c |} 
  \hline
  Shape & Field Function \\ \hline
  & \\
  & $1.0 / (\ln(\ln(\ln(exp(\sin(\ln(\cos(\theta)*(((d+0.761214)*\ln(d))\ +$ \\
   & $\ln(d-\theta))))*(\theta-e^\theta)))-((\theta-d)*(d/\theta)))-((\theta-e^\theta)\ *$ \\
   quarter-moon & $\ln(\ln(exp(\sin(\ln(\cos(\theta)*(((d+0.761214)*\ln(d))\ +$ \\
   & $\ln(d-\theta))))*(((\theta-e^\theta)-((\theta-e^\theta)*(d/\theta)))+ 0.432846)))\ -$ \\
   & $((\theta-d)*(d/\theta))))))$ \\
  & \\
  \hline
  & \\
  ellipse &  
  $1.0 / \cos(\theta) + \ln (d)$ \\
  & \\ \hline
  & \\
  discs &
  $1.0 / (\cos(\sin(\theta)-(\cos(\theta)-(\ln(-0.367378)/(\theta+d))))+\ln    (d))$ \\
  & \\ \hline
  & \\
  lines &
  $1.0 / (\ln(\ln(\ln((\cos(\cos(d))+(d*(3d+\ln(\theta)+\ln(\ln(\theta)))))+$ \\
  & $(((\ln(\theta)/(0.285192))*((1.0/\theta)+0.423969))/d))+d)+d))$ \\
  & \\ \hline
\end{tabular}
\end{center}
\caption{The field functions for the MPs utilized in this study. Note that exp(x) signifies $e^x$.}
\label{table:functions}
\end{table}
}

In order to identify the significant, distinguishing  moments
for each shape, aggregation simulations (usually several hundred) with
unbiased initial conditions are first performed.
The shapes of the final outcomes are visually inspected and placed
into categories.
For each final shape, the mean and standard deviation
of the four statistical moments of the evolving system are computed
over the entire shape aggregation process. 
For simulations starting with uniform random, unbiased initial conditions, 
the final outcomes of the quarter-moon,
ellipse and discs shapes evenly split into two categories,
with roughly $50\%$ of the final outcomes
belonging to each class. The outcomes of the parallel line shapes
are unbalanced, with
$84.1\%$ belonging to the majority class (two lines) and $15.9\%$
belonging to the minority class (one line).
These results, as well as the details that follow in the remainder
of the section, are summarized in Table~\ref{table:stats2}.

{\renewcommand{\arraystretch}{1.2}
\begin{table*}[t]
\begin{center}
\begin{tabular}{|c|c|c|c|}
\hline
\multicolumn{4}{|c|}{Quarter-moon} \\ \hline
Shapes & \parbox{2cm}{\vspace{1mm}Unbiased Percentage\vspace{1mm}} & Thresholded Moment & \parbox{1.9cm}{Biased\\ Percentage}
\\ \hline
left-pointing & $50\%$ & $Skewness_x \ge 0.315$  &  $100\%$  
\\ \hline
right-pointing & $50\%$ & $Skewness_x \le -0.315$  &  $100\%$ 
\\ \hline
\hline
\multicolumn{4}{|c|}{Ellipse} \\ \hline
Shapes & \parbox{2cm}{\vspace{1mm}Unbiased Percentage\vspace{1mm}} & Thresholded Moment & \parbox{1.9cm}{Biased\\ Percentage}
\\ \hline
single ellipse & $50\%$ & $Kurtosis_y \ge 2.18$  &  $100\%$ 
\\ \hline
non-single ellipse & $50\%$ &  & $0\%$
\\ \hline
two ellipses & $-$ & $Kurtosis_y \le 1.85$  & $100\%$
\\ \hline
\hline
\multicolumn{4}{|c|}{Multiple Discs} \\ \hline
Shapes & \parbox{2cm}{\vspace{1mm}Unbiased Percentage\vspace{1mm}} & Thresholded Moment & \parbox{1.9cm}{Biased\\ Percentage}
\\ \hline
three discs & $0\%$ & $Variance_x \le 10,270$  &  $100\%$ 
\\ \hline
no three discs & $100\%$ &$Variance_x \ge 15,500$ & $100\%$
\\ \hline
Shapes & \parbox{2cm}{\vspace{1mm}Unbiased Percentage\vspace{1mm}} & Thresholded Moment & \parbox{1.9cm}{Biased\\ Percentage}
\\ \hline
three discs & $0\%$ &  &  $4\%$ 
\\ \hline
four discs & $50\%$ & $1.99 \le Kurtosis_x \le 2.09$  & $75\%$
\\ \hline
five or more discs & $50\%$ & & $21\%$
\\ \hline
\hline
\multicolumn{4}{|c|}{Parallel Line Segments} \\ \hline
Shapes & \parbox{2cm}{\vspace{1mm}Unbiased Percentage\vspace{1mm}} & Thresholded Moment & \parbox{1.9cm}{Biased\\ Percentage}
\\ \hline
two lines & $ 84.1\%$ & $Kurtosis_x \le 1.90$ &  $100\%$  
\\ \hline
one line  & $15.9\%$ & $Kurtosis_x \ge 2.29$ &  $100\%$ 
\\ \hline   
\end{tabular}
\end{center}
\caption{Table summarizing results generated with unbiased and biased
initial conditions.}
\label{table:stats2}
\end{table*}
}

\begin{figure}
\begin{center}
\includegraphics[width=0.9\linewidth]{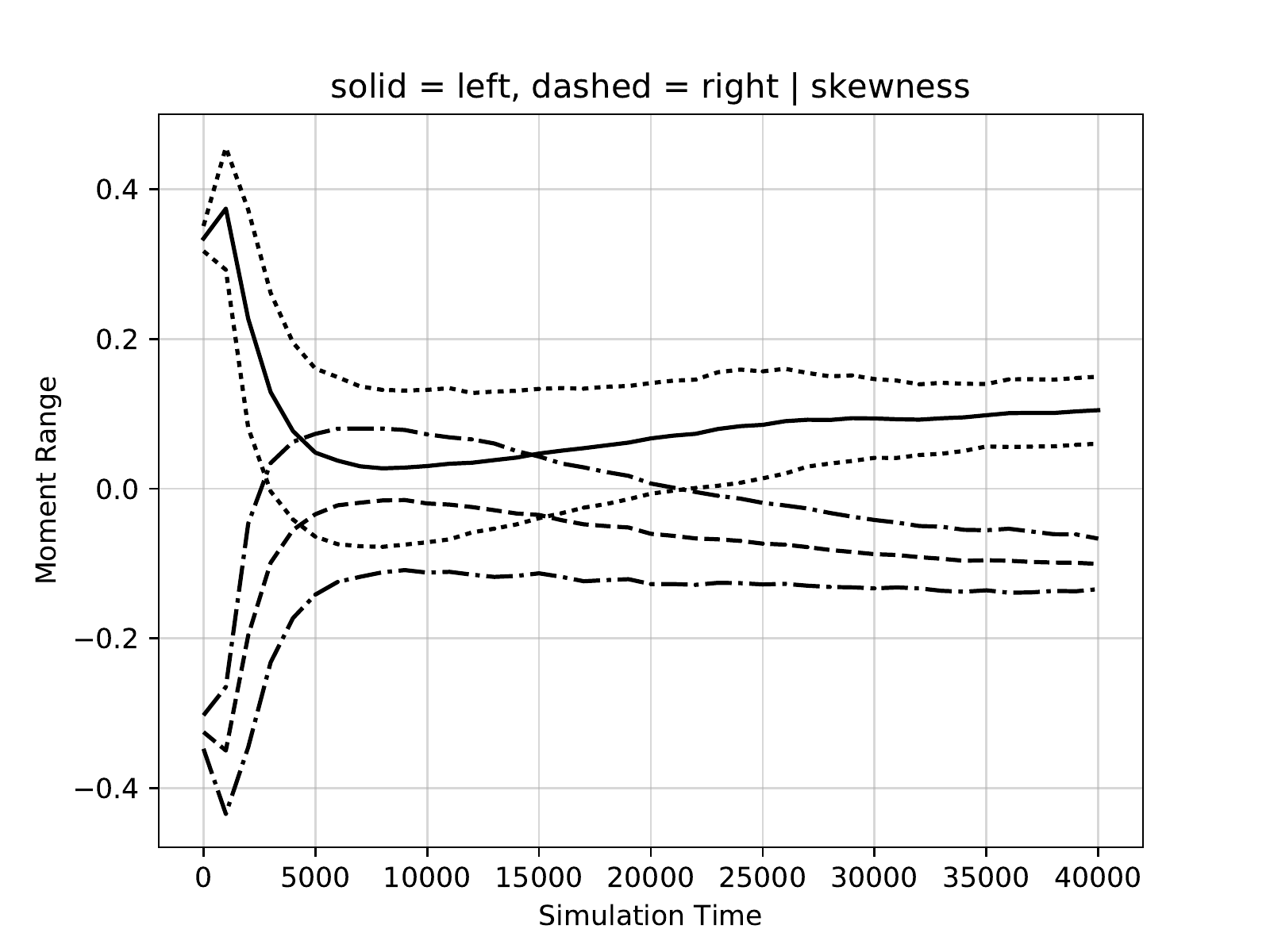}
\end{center}
\caption{Skewness of the $x$ coordinate of the biased quarter-moon shape
aggregations over time.}
\label{fig:quartermoon_x3_biased}
\end{figure}

A typical unbiased aggregation for the quarter-moon shape
is shown in Figure~\ref{fig:quartermoon}.
The simulation reaches a
stable state by 35,000 simulation steps. 
We identified skewness in the $x$ coordinate to be the significant
macroscopic feature for this shape.
See Figure~\ref{fig:quartermoon_x3}
for the evolution of this feature over the course of the aggregation
given uniform random initial conditions.
Two sets of
biased initial conditions (each with 100 examples) were generated
with constrained skewness values in the $x$ coordinate, with the thresholds
set as greater than $0.315$ (2 standard deviations from the mean) and
less than $-0.315$.
Since the distributions are generated stochastically, they
do not have the exact targeted skewness value. So our acceptance test
is based on a threshold.
We performed simulations with the quarter-moon interaction
function for these $200$ biased initial conditions. Of the $100$ initial
conditions with a thresholded third $x$ moment below $-0.315$, $100\%$ of
the final outcomes are right-pointing structures. 
Of the $100$ initial conditions with a thresholded third $x$ moment above 
$0.315$, $100\%$ are left-pointing structures.
Figure~\ref{fig:quartermoon_x3_biased}
presents the evolution of this feature over the course of the aggregation
given the biased initial conditions.

\begin{figure}
\begin{center}
\includegraphics[width=1.0\linewidth]{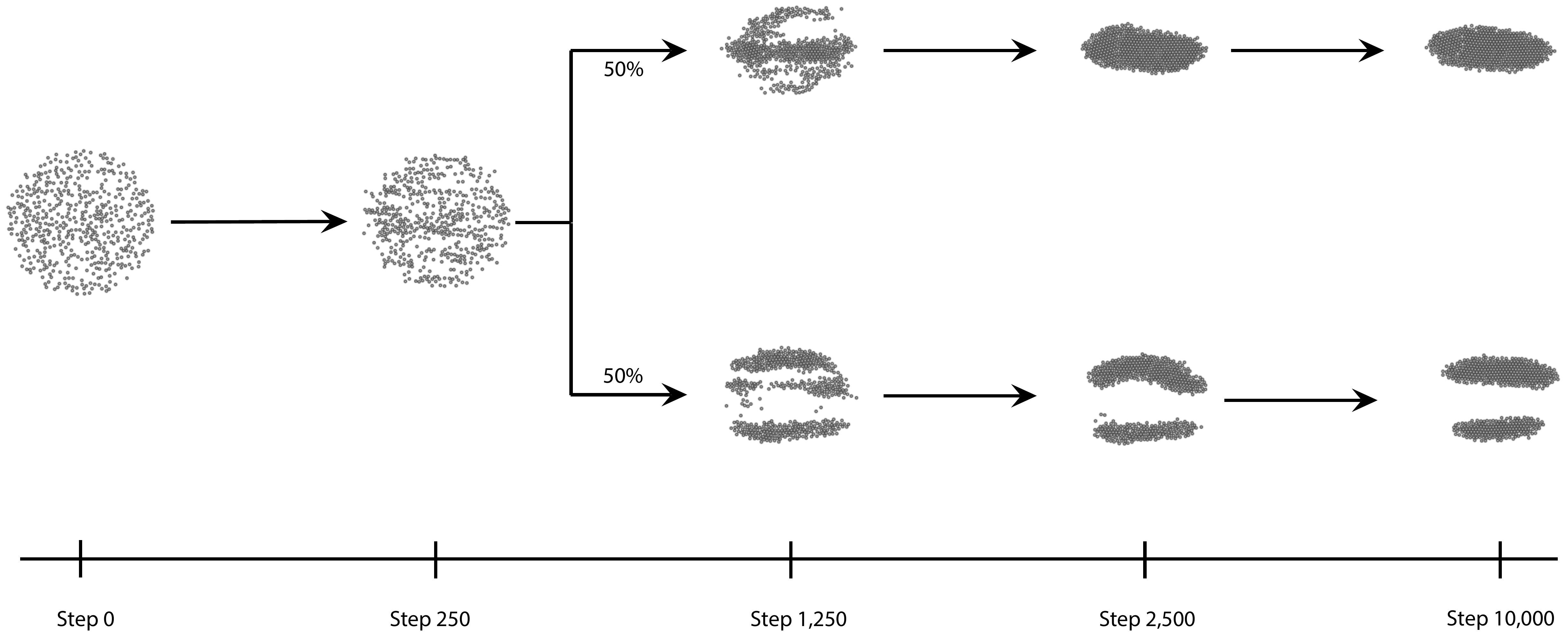}
\end{center}
\caption{Shape aggregation of the ellipse MPs starting from random
initial conditions.}
\label{fig:ellipse}
\end{figure}

\begin{figure}[p]
\begin{center}
\includegraphics[width=0.9\linewidth]{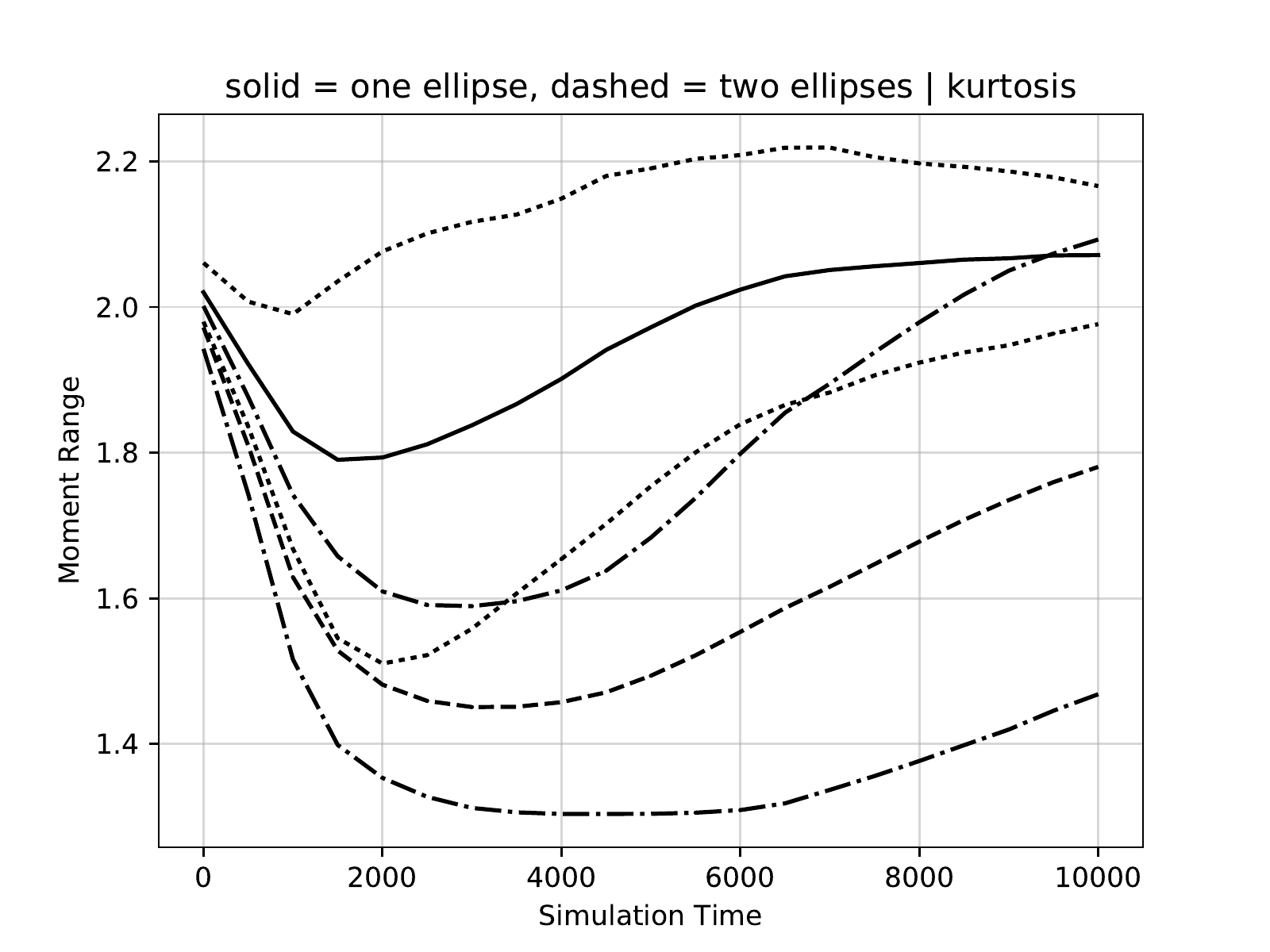}
\end{center}
\caption{Kurtosis of the $y$ coordinate of the unbiased ellipse shape
aggregations over time.}
\label{fig:ellipse_y4}
%
\begin{center}
\includegraphics[width=0.9\linewidth]{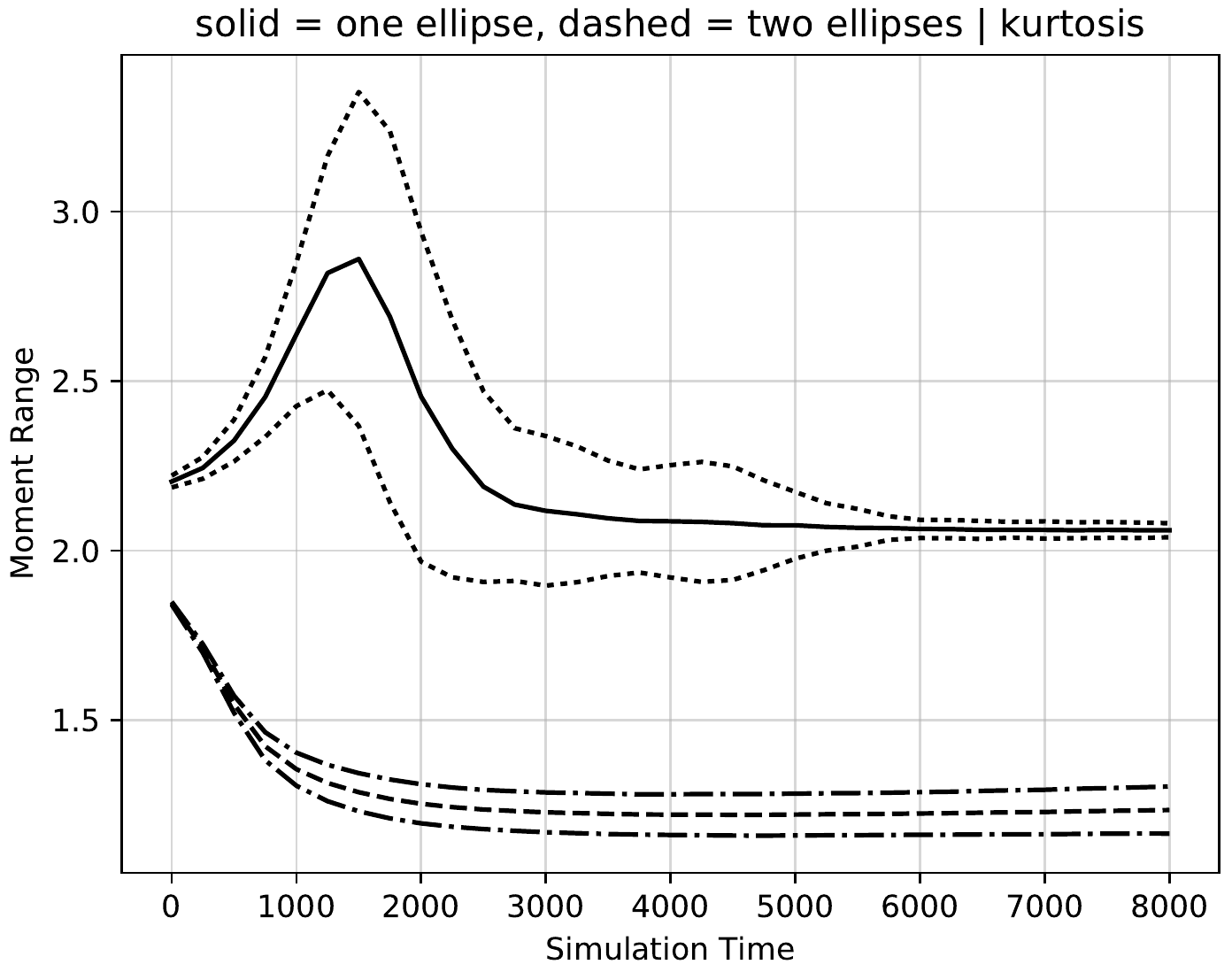}
\end{center}
\caption{Kurtosis of the $y$ coordinate of the biased ellipse shape
aggregations over time.}
\label{fig:ellipse_y4_biased}
\end{figure}

A typical unbiased aggregation for the ellipse shape
is shown in Figure~\ref{fig:ellipse}, with half of the unbiased
initial conditions producing a single ``perfect'' ellipse, with the
other half producing either two ellipses or a deformed ``blob''.
The unbiased simulation is computed for 10,000 steps.  If the 
simulations are run for 50,000 steps they all will produce a single
ellipse.
Kurtosis in the $y$ direction was found to 
be the significant macroscopic feature for this shape.
See Figure~\ref{fig:ellipse_y4}
for the evolution of this feature over the course of the aggregation
given uniform random initial conditions.
100 simulations 
were performed with initial conditions that had their $y$ coordinates' 
kurtosis thresholded to be above $2.18$.
Given these biased initial conditions all 
simulation ($100\%$) produced a perfect single ellipse by step 7,500.
100 additional simulation were performed with initial conditions that
had their $y$ kurtosis set below 1.85. 
All 100 simulations produced two ellipses by step 7,500.
Figure~\ref{fig:ellipse_y4_biased}
presents the evolution of this feature over the course of the aggregation
given the biased initial conditions. These results show that not only 
can biased initial conditions direct the outcomes of the simulations,
but
they can also significantly speed up the formation of the desired result, 
with the single ellipse being guaranteed to form by 50,000 steps in the
unbiased case and by 7,500 steps given biased initial conditions.

\begin{figure}[p]
\begin{center}
\includegraphics[width=1.0\linewidth]{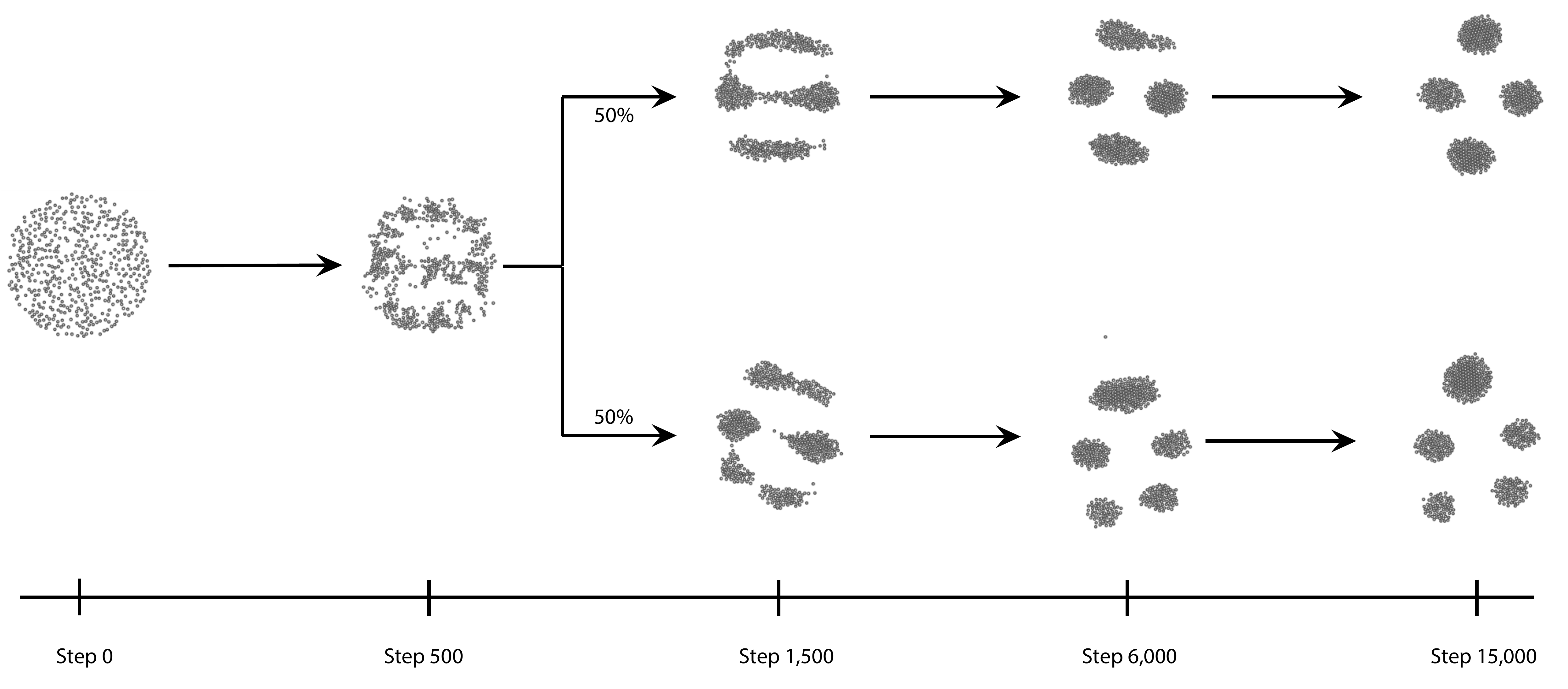}
\end{center}
\caption{Shape aggregation of the discs MPs starting with random
initial conditions.}
\label{fig:blobs}

\end{figure}

\begin{figure}
\begin{center}
\includegraphics[width=0.9\linewidth]{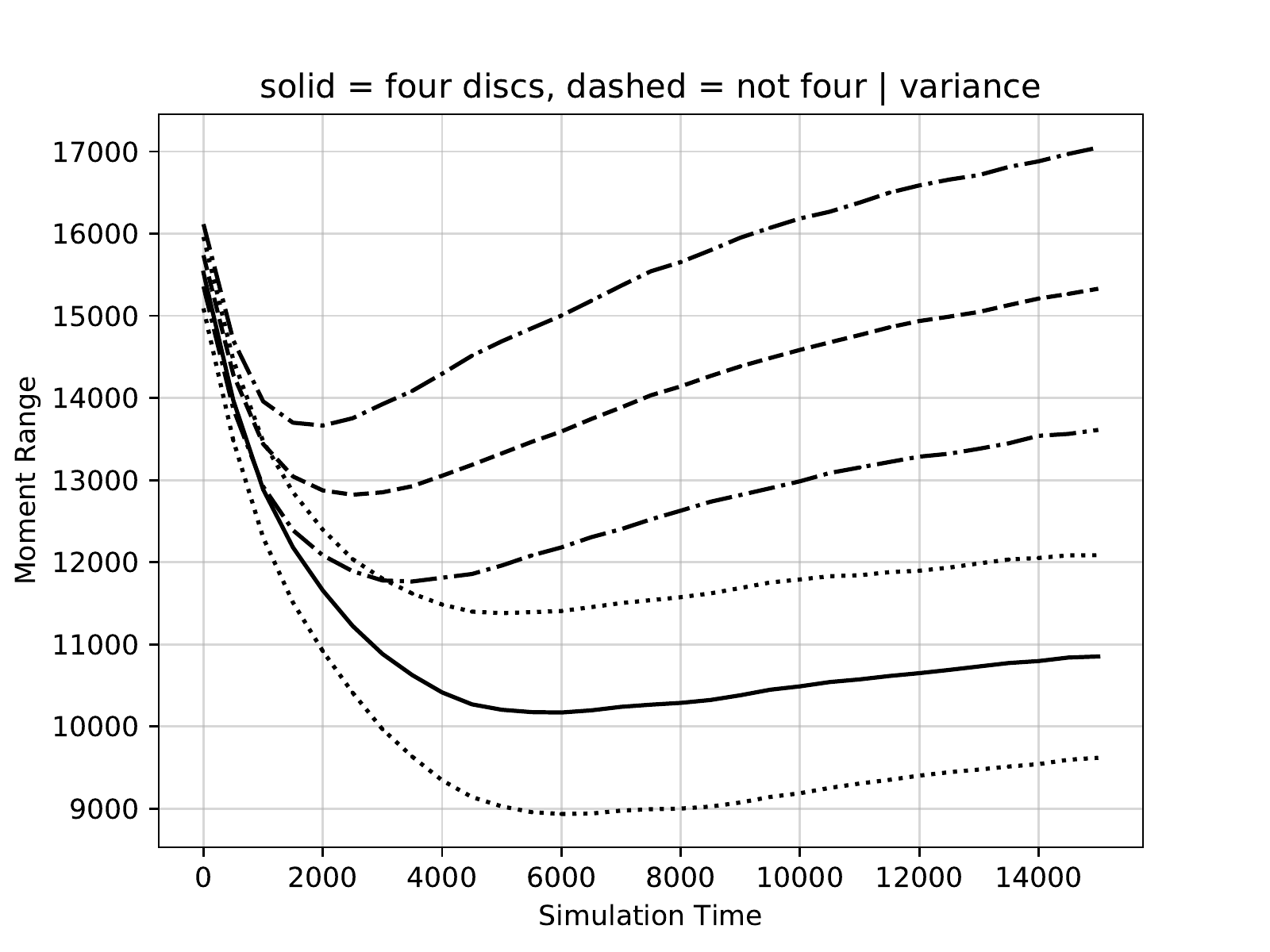}
\end{center}
\vspace{-3mm}
\caption{Variance of the $x$ coordinate of the unbiased discs shape
aggregations over time.}
\label{fig:blobs_x2}
\begin{center}
\includegraphics[width=0.9\linewidth]{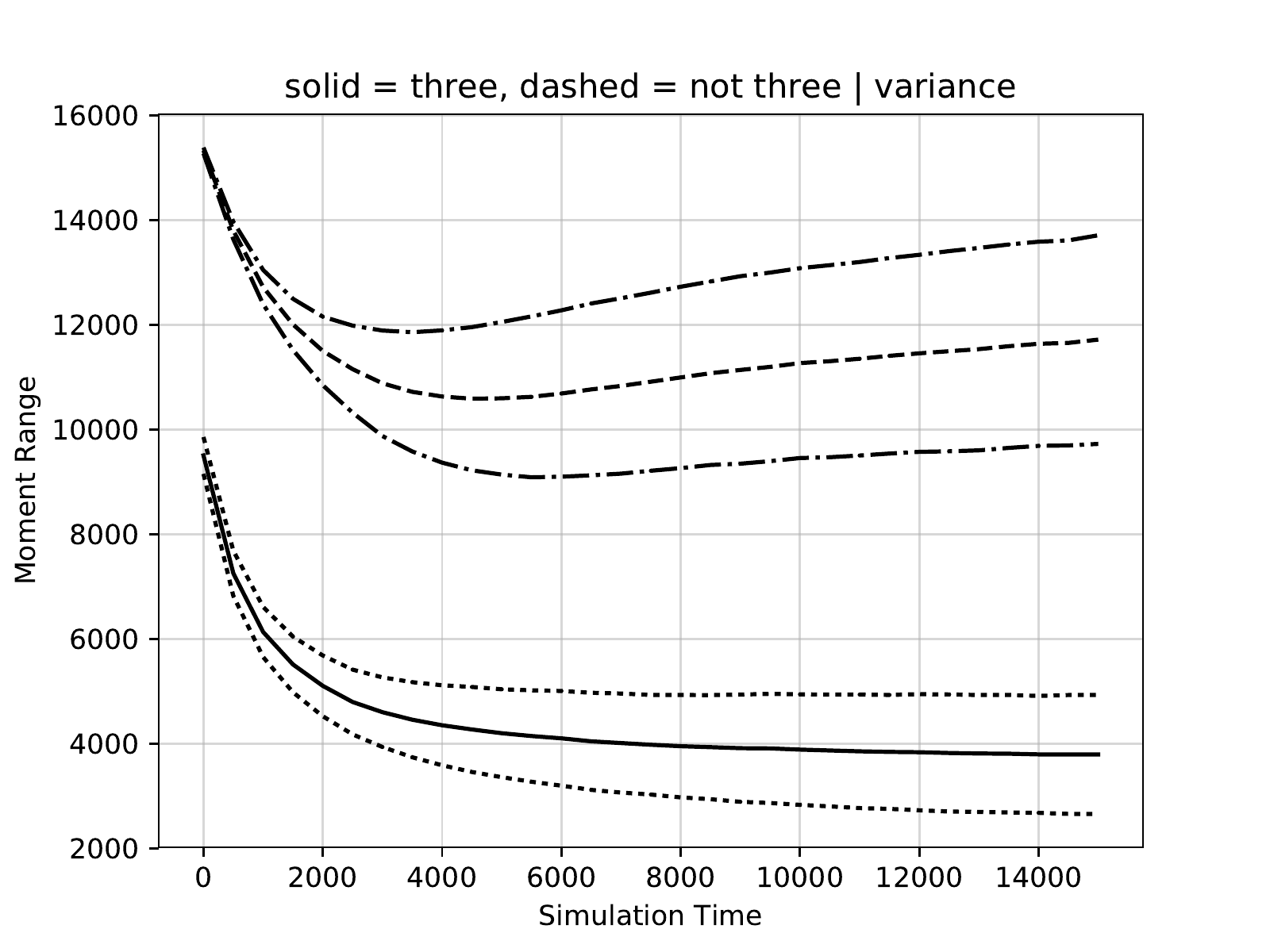}
\end{center}
\vspace{-3mm}
\caption{Variance of the $x$ coordinate of the biased discs shape
aggregations over time.}
\label{fig:blobs_x2_biased}
\end{figure}

The discs dataset, when run with 200 unbiased initial conditions,
produces $100$ four discs structures and $100$
structures of five or more discs, with a typical shape aggregation shown in
Figure~\ref{fig:blobs}. The simulation reaches a
stable state by 15,000 steps. 
We identified variance in the $x$ direction to be the significant
macroscopic feature for this shape. See Figure~\ref{fig:blobs_x2}.
In our experiments as the variance of the initial conditions was
lowered, we found that we could generate a new shape (one that did
not appear with unbiased initial conditions), that contained
only three discs.
Thresholding the $x$ variance of the initial conditions
to be less than 10,270 would always ($100\%$) produce a 3-disc result.
A typical 3-disc shape  is  presented in
Figure \ref{fig:biased_results}(c).
We were unable to consistently generate a 4-disc
result for most simulations by thresholding the variance.
Thresholding the $x$ kurtosis to be greater than 1.90 and less than 2.09
did lead to an increased
number of 4-disc results ($75\%$), which we deemed as less than
consistent or robust.
Figure \ref{fig:blobs_x2_biased} presents
the evolution of $x$ variance over the course of the 3-disc aggregation
given the biased initial conditions. Note that 
no 3-disc results were produced when keeping the $x$ variance above
15,500 in the biased initial conditions.
 
\begin{figure}[t]
\begin{center}
\includegraphics[width=1.0\linewidth]{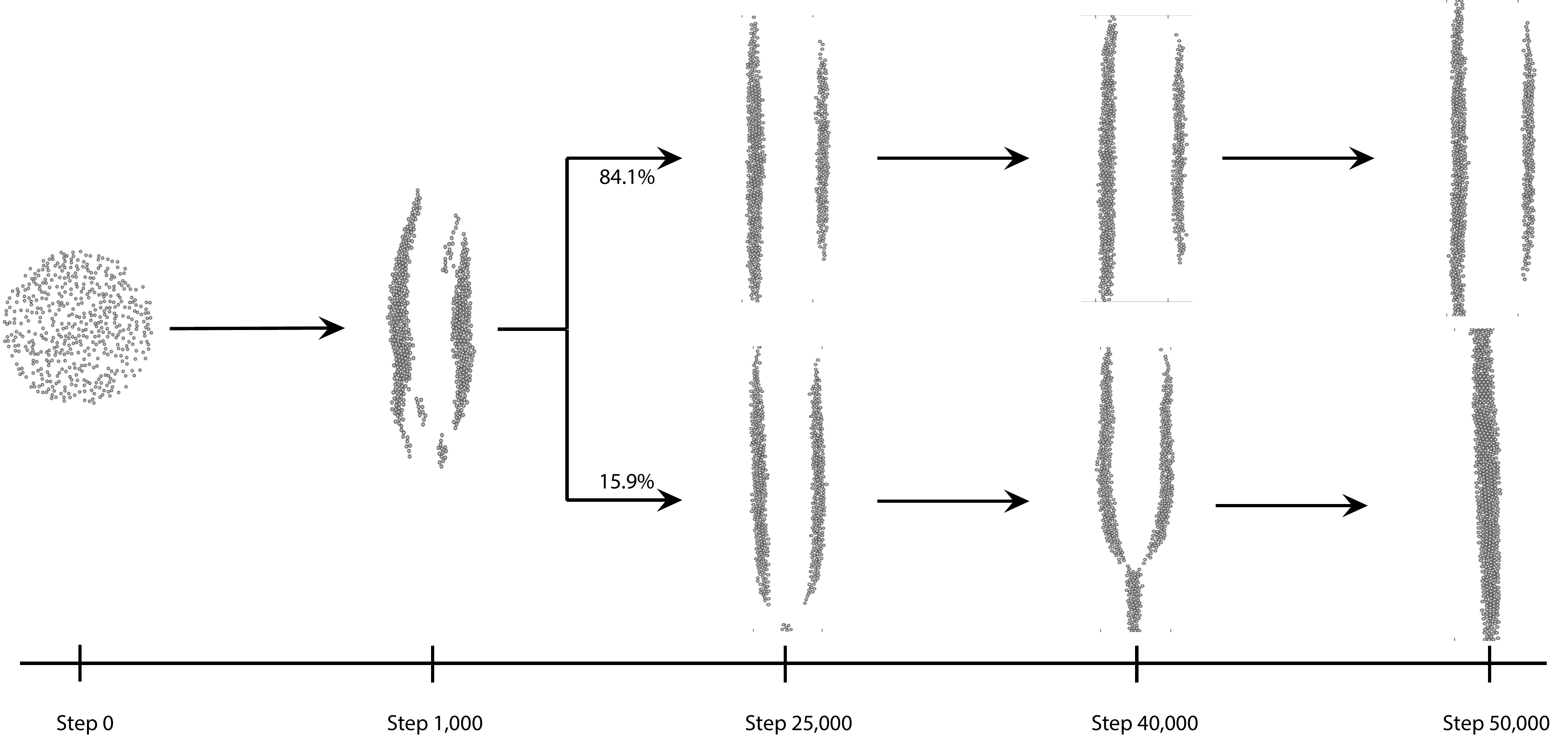}
\end{center}
\caption{Shape aggregation of the line segment MPs starting from
random initial conditions.}
\label{fig:vlines}
\end{figure}

\begin{figure}[p]
\vspace{4mm}
\begin{center}
\includegraphics[width=0.9\linewidth]{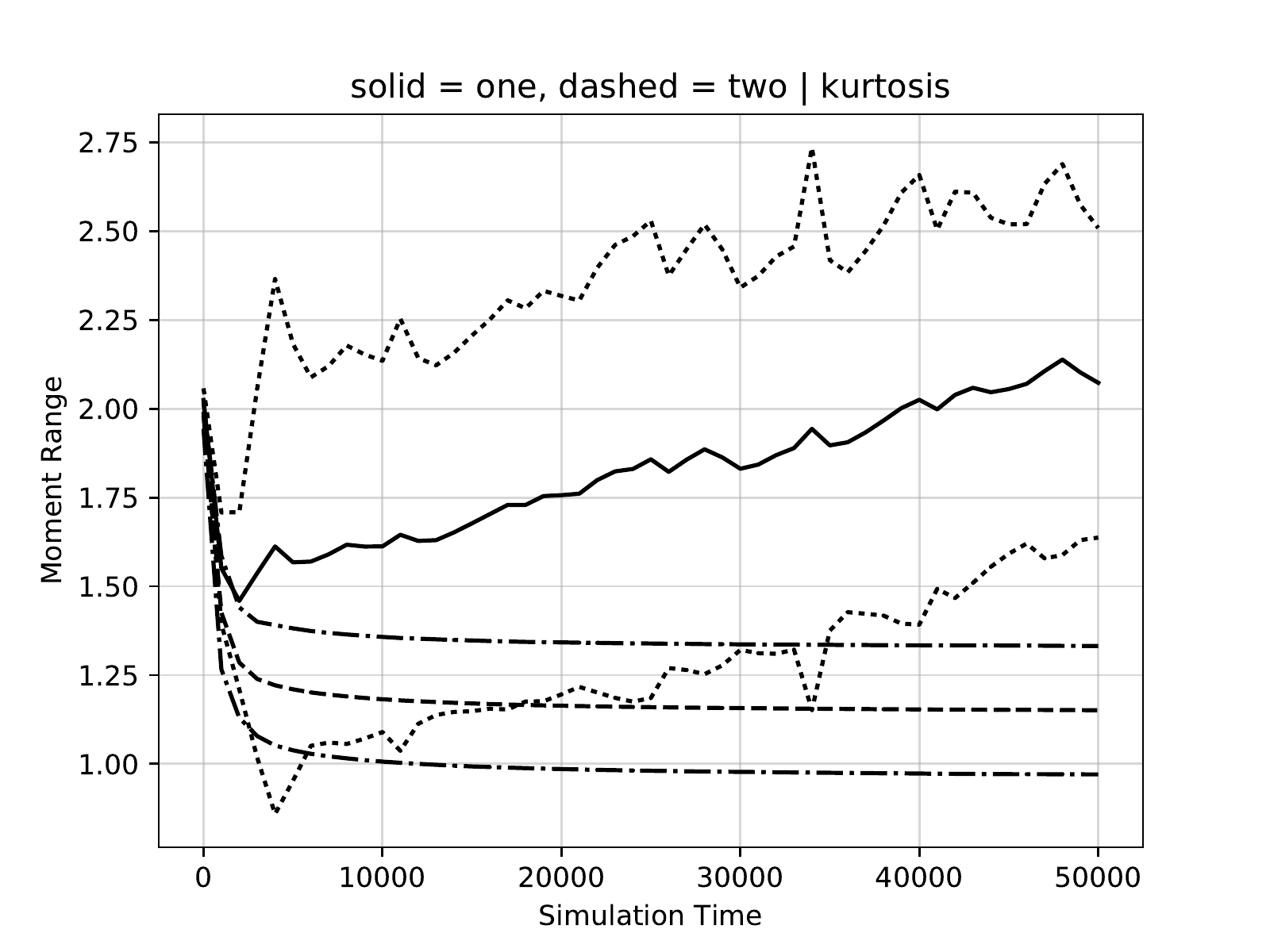}
\end{center}
\caption{Kurtosis of the $x$ coordinate of the unbiased line segment shape
aggregations over time.}
\label{fig:vlines_x4}
%
\vspace{4mm}
\begin{center}
\includegraphics[width=0.9\linewidth]{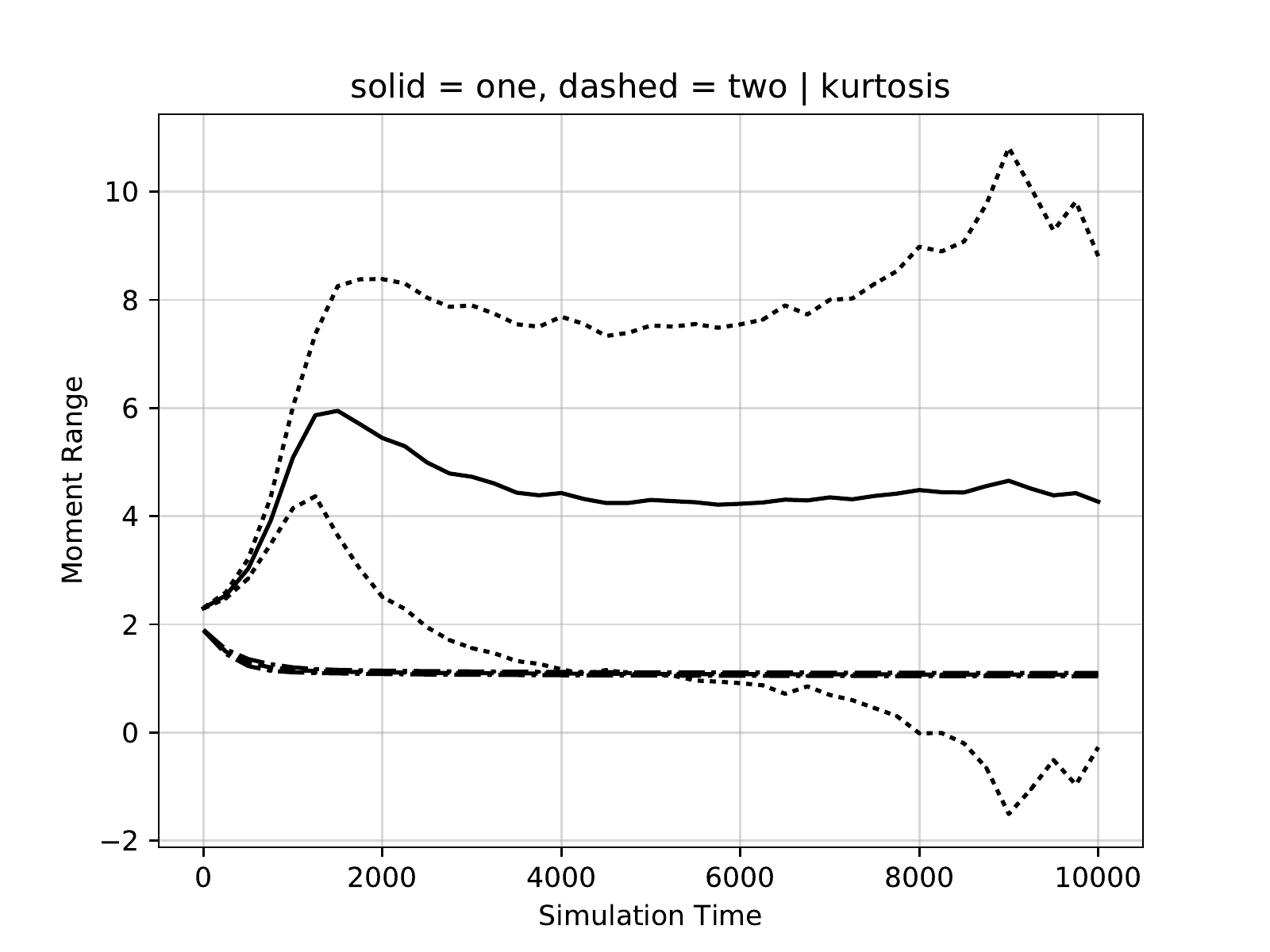}
\end{center}
\caption{Kurtosis of the $x$ coordinate of the biased line segment shape
aggregations over time.}
\label{fig:vlines_x4_biased}
\end{figure}

The parallel line dataset, when run with unbiased initial conditions,
contains $526$ instances of two vertical parallel
line segments ($84.1\%$) and $100$ instances of 
one vertical line ($15.9\%$), as seen in
Figure~\ref{fig:vlines}. The simulation
reaches a stable state by 50,000 steps. 
We identified kurtosis in the $x$ coordinate to be the significant
macroscopic feature for the shape.  See Figure~\ref{fig:vlines_x4}.
100 simulations were performed with initial conditions that had their
$x$ coordinates' kurtosis thresholded to be below 1.90. Given these biased
initial conditions all simulation (100\%) produced the two line structure.
100 simulations were then performed with initial conditions that had their
$x$ coordinates' kurtosis thresholded to be above 2.29. These biased
conditions produced results that consistently ($100\%$)
created the minority class structure of a single line by 10,000 steps.
Figure \ref{fig:vlines_x4_biased} presents
the evolution of $x$ kurtosis over the course of the aggregation
given the biased initial conditions.

\section{Discussion}
Our experiments show that our agents (MPs) can be reliably directed
to form into large-scale, macroscopic structures using local-only
behaviors, based on chemical diffusion fields and 
biased initial conditions; thus producing a stigmergic
phenomenon \cite{Theraulaz99}.
This type of global outcome is of particular 
interest to those looking for a robust, adaptable, and independent 
self-organizing system. A review conducted by the European Space Agency
has shown that for their harsh 
working environment (space, Mars, etc...) the robustness of a 
local-only system is a key consideration and would allow for 
relatively simple (and easy to transport) satellites/equipment to 
form into a larger and more complex system that would be impossible 
to transport as a monolithic structure \cite{spaceWeb}.
Systems using global information, with centralized communication 
between primitives, or a command-and-control structure, can
form more complex shapes more quickly than local-only approaches.
However these
systems can fail if this communication is interrupted
or the command-and-control structure breaks down \cite{Pettazzi2006SNR}. 

The reasons that certain biased initial conditions may be used to 
direct the outcome of an aggregation process are frequently visually 
obvious. For example, the biased initial conditions seen in Figure
\ref{fig:biased_results}(a) are clearly skewed to the right side
of the arena. So it is clear that the majority of the agents are
already amassed around the center of the object to be formed. This
can also be seen in Figure \ref{fig:biased_results}(c), where lowering
the variance of the X component makes the initial distribution of the
MPs cluster around the central axis which the three discs will
form along. The visual evidence that biasing initial conditions
effectively pre-starts the aggregation process towards a desired shape is
not as evident in the examples that constrain kurtosis ((b) and (d)).
A higher Y kurtosis value in example (b) means that there should be a
higher concentration of agents along the $Y=500$ axis. In (d), a lower
X kurtosis value means that
there should be more agents distributed away from the center of the arena.
But in both of these cases, while statistically this is true, it is not
visually evident.


It should be noted that not all shapes are stable throughout the
simulation. The
vertical lines being a good example; up to approximately 5,000 steps the
1-line and 2-lines shapes look very similar and
have similar statistical moments, however between 10,000 and 45,000 steps
$15.9\%$ of the simulations will converge into a single line. This shows
that some shapes may appear to be stable at one point during a simulation,
when actually they have not yet stabilized. 
Or in other words, our agents
may produce more than one type of distinct shape during their evolution.
Thus, our experiments show that we are able to consistently produce certain
self-organized swarms at a specific point in time during their aggregation.


For these types of self-organizing systems, it is clearly desired to
have a completely local solution. Given that the methods described
here require the computation of global system-level quantities (moments of
the entire distribution) and the manipulation of the locations of the
agents in order to meet some global constraint, we have not achieved
this goal. In the future, we intend to explore if our genetic programming
approach to local chemical field evolution, that leads to the formation
of macroscopic shapes, can also be used to find chemical fields (i.e.
local interactions) that direct a swarm that is uniformly randomly
distributed into one that has specific statistical properties. This
would lead to a two-step approach that is truly based on local-only
interactions. In this case, the agents, which have been uniformly
randomly placed in an environment, would follow chemical fields that
move them into a biased initial condition. Then they would switch
to a field that robustly directs them to form in a specific macroscopic
shape.

Additional future work with this system will involve robustness testing of
the parameters of the initial conditions. Previous work in this area has
defined a lower-bound
for the number of primitives required to successfully create a single
ellipse (400), but has not defined an upper-bound \cite{Bai08c}. An
upper and 
lower bound on the number of primitives (which of course is related to
MP density) in the simulation is important as 
more complex shapes are created. Additionally future work will involve 
further analysis of the aggregation processes that cannot be completely 
controlled. These investigations should reveal new features that 
differentiate swarms that can be controlled via the reported method and 
those that cannot. We imagine that manipulation of other features will 
further enhance our ability to direct our self-organizing system.

Ultimately we would like to implement our spatial self-organization
approach in a swarm robotics system.  We believe that this would be
feasible with robots that communicate locally via Bluetooth. Given that
all robots emit the same field, a single robot can compute the cumulative
chemical field and its gradient at its location simply by knowing the
distances and angular relationships of neighboring robots from itself.
If each robot has a unique broadcasted
ID, we imagine that distance could be derived from signal strength and
angular information could be computed from inputs from multiple
antennae. We have had discussions with a local roboticist about
the possibility of using her swarm robotics platform to investigate
our methods for distributed control.


\section{Conclusions}

We have previously developed an agent-based self-organizing shape formation
system. The agents perform identical behaviors based on sensing local
information emitted into the environment by the agents.
Genetic programming may be used to discover local interaction rules that
lead the agents to self-organize into a number of user-specified shapes.
However, since the agents are initially uniformly randomly 
placed in the environment and they stochastically follow prescribed rules,
the aggregation simulations do not always produce the same final results.
In order to develop methods that could be used to direct the agents to
robustly form one specific configuration, we explored the relationships
between an agent swarm's moments and its final configuration.
After having shown that these moments could be used to predict the
outcome of an MP aggregation in previous work, we demonstrate in this
work that biasing the swarm's initial conditions based on these moments
can be used to consistently direct the swarm to produce a desired
macroscopic shape.
 
By analyzing the statistical moments of the agents' positions over the
entire shape aggregation process, we have identified significant,
distinguishing moment features, and utilize them as constraints on
simulation initial conditions for a number of bifurcating shapes. 
Biased initial conditions may be generated that meet these moment
constraints, which then affect the resulting shape outcomes.
In almost all of our examples we can completely control the result
of the self-organization process. In some other cases we can significantly
increase the likelihood of producing a desired configuration.
In a more general sense, our work also indicates that complex,
non-linear dynamical self-organizing systems may be controlled by
manipulating their initial conditions.


\section*{Disclosure/Conflict-of-Interest Statement}
%
 The authors declare that the research was conducted in the absence of any commercial or financial relationships that could be construed as a potential conflict of interest.

\section*{Acknowledgments}
The authors would like to thank
Robert Gilmore, Christian Kuehn and Santiago Onta\~{n}\'{o}n
for many helpful discussions and suggestions.
This research was funded by
National Science Foundation grants CCF-0636323 and IIS-0845415.

\bibliographystyle{gPAA}
\bibliography{IJPEDS17}

\newpage

\section*{Figures}


\fboxsep=0mm 
\fboxrule=1.5pt 

\clearpage

\clearpage

\clearpage

\clearpage

\clearpage

\clearpage

\clearpage

\clearpage

\end{document}